\documentclass[11pt]{article}
\usepackage{amssymb}
\begin{document}
\renewcommand{\Bbb}{\mathbb}

%\batchmode

\thispagestyle{empty}

%\input{psfig}

%%%%%%%%%%%%%%%%%%%%%%%%%%%%%%%%%%%%%%%%%%%%%%%%%%%%%%%%%%%%%%%%%%%%%%%%%
%                            GREEK                                      %
%%%%%%%%%%%%%%%%%%%%%%%%%%%%%%%%%%%%%%%%%%%%%%%%%%%%%%%%%%%%%%%%%%%%%%%%%
\newcommand{\al}{\alpha}
\newcommand{\bet}{\beta}
\newcommand{\ga}{\gamma}
\newcommand{\del}{\delta}
\newcommand{\ep}{\epsilon}
\newcommand{\epx}{\varepsilon}
\newcommand{\ze}{\zeta}
\renewcommand{\th}{\theta}
\newcommand{\thx}{\vartheta}
\newcommand{\io}{\iota}
\newcommand{\la}{\lambda}
\newcommand{\ka}{\kappa}
\newcommand{\pix}{\varpi}
\newcommand{\rhx}{\varrho}
\newcommand{\si}{\sigma}
\newcommand{\six}{\varsigma}
\newcommand{\yp}{\upsilon}
\newcommand{\om}{\omega}
\newcommand{\phx}{\varphi}
\newcommand{\Ga}{\Gamma}
\newcommand{\De}{\Delta}
\newcommand{\Th}{\Theta}
\newcommand{\La}{\Lambda}
\newcommand{\Si}{\Sigma}
\newcommand{\Yp}{\Upsilon}
\newcommand{\Om}{\Omega}

%%%%%%%%%%%%%%%%%%%%%%%%%%%%%%%%%%%%%%%%%%%%%%%%%%%%%%%%%%%%%%%%%%%%%%%%%
%                         MATH MODE COMMAND                             %
%%%%%%%%%%%%%%%%%%%%%%%%%%%%%%%%%%%%%%%%%%%%%%%%%%%%%%%%%%%%%%%%%%%%%%%%%
\renewcommand{\L}{\cal{L}}
\newcommand{\M}{\cal{M}}
\newcommand{\be}{\begin{eqnarray}}
\newcommand{\ee}{\end{eqnarray}}
\newcommand{\jt}{\tilde{J}}
\newcommand{\Ra}{\Rightarrow}
\newcommand{\lra}{\longrightarrow}
\newcommand{\ti}{\tilde}
\newcommand{\pj}{\prod J}
\newcommand{\pjt}{\prod\tilde{J}}
\newcommand{\prb}{\prod b}
\newcommand{\prc}{\prod c}
\newcommand{\bft}{|\tilde{\phi}>}
\newcommand{\bfj}{|\phi>}
\newcommand{\lan}{\langle}
\newcommand{\ran}{\rangle}
\newcommand{\bz}{\bar{z}}
\newcommand{\bJ}{\bar{J}}
\newcommand{\vacr}{|0\rangle}
\newcommand{\vacl}{\langle 0|}
\newcommand{\IFF}{\Longleftrightarrow}
\newcommand{\phr}{|phys\ran}
\newcommand{\phl}{\lan phys|}
\newcommand{\nonu}{\nonumber\\}
\newcommand{\tg}{\tilde{g}}
\newcommand{\tM}{\ti{M}}
\newcommand{\hd}{\hat{d}}
\newcommand{\hL}{\hat{L}}
\newcommand{\sir}{\si^\rho}
\newcommand{\mf}{\mathfrak}
\newcommand{\mfg}{{\mathfrak{g}}}
\newcommand{\mfbg}{{\bar{\mathfrak{g}}}}
\newcommand{\mfk}{{\mathfrak{k}}}
\newcommand{\mfc}{{\mathfrak{c}}}
\newcommand{\mfbc}{{\bar{\mathfrak{c}}}}
\newcommand{\mfbk}{{\bar{\mathfrak{k}}}}
\newcommand{\mfn}{{\mathfrak{n}}}
\newcommand{\mfbn}{{\bar{\mathfrak{n}}}}
\newcommand{\mfh}{{\mathfrak{h}}}
\newcommand{\mfbh}{{\bar{\mathfrak{h}}}}
\newcommand{\mfp}{{\mathfrak{p}}}
\newcommand{\mfbp}{{\bar{\mathfrak{p}}}}
\newcommand{\mfU}{{\mf U}}
\newcommand{\Lg}{\L^{(\mfg})}
\newcommand{\Lgk}{\L^{(\mfg,\mfk)}}
\newcommand{\Lgkp}{\L^{(\mfg,\mfk')}}
\newcommand{\Lk}{\L^{(\mfk)}}
\newcommand{\Lkp}{\L^{(\mfk')}}
\newcommand{\Lc}{\L^{(\mfc)}}
\newcommand{\Lgg}{\L^{(\mfg,\mfg)}(\la)}
\newcommand{\Lgp}{\L^{(\mfg,\mfg')}(\la)}
\newcommand{\Mgp}{\M^{(\mfg,\mfg')}(\la)}
\newcommand{\Mgk}{\M^{(\mfg,\mfk)}(\la)}
\newcommand{\Hgk}{H^{(\mfg,\mfk)}}
\newcommand{\Mgkp}{\M^{(\mfg,\mfk^\prime)}}
\newcommand{\Hgp}{H^{(\mfg,\mfg')}}
\newcommand{\vo}{v_{0\la}}
\newcommand{\Ml}{\M(\la)}
\newcommand{\Ll}{\L(\la)}
\newcommand{\RR}{\mathbb{R}}
%%%%%%%%%%%%%%%%%%%%%%%%%%%%%%%%%%%%%%%%%%%%%%%%%%%%%%%%%%%%%%%%%%%%%%%%%
%                         MISCELLANEOUS                                 %
%%%%%%%%%%%%%%%%%%%%%%%%%%%%%%%%%%%%%%%%%%%%%%%%%%%%%%%%%%%%%%%%%%%%%%%%%

\newcommand{\e}[1]{\label{e:#1}\end{eqnarray}}
 \renewcommand{\r}[1]{(\ref{e:#1})}
 \newcommand{\ind}{\indent}
\newcommand{\np}{\newpage}
\newcommand{\hs}{\hspace*}
\newcommand{\vs}{\vspace*}
\newcommand{\nl}{\newline}
\newcommand{\bqu}{\begin{quotation}}
\newcommand{\equ}{\end{quotation}}
\newcommand{\bit}{\begin{itemize}}
\newcommand{\eit}{\end{itemize}}
\newcommand{\ben}{\begin{enumerate}}
\newcommand{\een}{\end{enumerate}}
\newcommand{\ul}{\underline}
\newcommand{\nn}{\nonumber}
\newcommand{\lef}{\left}
\newcommand{\rig}{\right}
\newcommand{\fra}{\twelvefrakh}
\newcommand{\Bb}{\twelvemsb}
\newcommand{\bT}{\bar{T}(\bz)}
\newcommand{\dagg}{^{\dagger}}
\newcommand{\qd}{\dot{q}}
\newcommand{\cP}{{\cal P}}
\newcommand{\hg}{\hat{g}}
\newcommand{\hh}{\hat{h}}
\newcommand{\hpg}{\hat{g}^\prime}
\newcommand{\htg}{\tilde{\hat{g}}^\prime}
\newcommand{\pri}{\prime}
\newcommand{\bis}{{\prime\prime}}
\newcommand{\lap}{\la^\prime}
\newcommand{\rhop}{\rho^\prime}
\newcommand{\Dgp}{\Delta_{g^\prime}^+}
\newcommand{\Dg}{\Delta_g^+}
\newcommand{\Pro}{\prod_{n=1}^\infty (1-q^n)}
\newcommand{\Pg}{P^+_{\hg}}
\newcommand{\Pgp}{P^+_{\hg\pri}}
\newcommand{\hmu}{\hat{\mu}}
\newcommand{\hnu}{\hat{\nu}}
\newcommand{\hrho}{\hat{\rho}}
\newcommand{\gp}{g^\prime}
\newcommand{\pp}{\prime\prime}
\newcommand{\CM}{\hat{C}(g',M')}
\newcommand{\CI}{\hat{C}(g',M^{\prime (1)})}
\newcommand{\CL}{\hat{C}(g',L')}
\newcommand{\HL}{\hat{H}^p (g',L')}
\newcommand{\HMI}{\hat{H}^{p+1}(g',M^{\prime (1)})}
\newcommand{\da}{\dagger}

\newcommand{\wro}{w^\rho}
\renewcommand{\Box}{\rule{2mm}{2mm}}
\begin{flushright}

%KAU-NT-FR-99/XX-SE\\
\end{flushright}
\vs{10mm}

\begin{center}

{\Large{\bf Equivalence of Chern-Simons gauge theory and WZNW model using a
BRST symmetry}}
\\
\vspace{10 mm}
{\large{Jens Fjelstad \footnote{email: jens.fjelstad@kau.se}\\ and\\
Stephen Hwang \footnote{email: stephen.hwang@kau.se}}
\vspace{4mm}\\
Department of Physics,\\ Karlstad
University, S-651 88 Karlstad, Sweden}

\vs{15mm}

{\bf Abstract }\end{center} 
\begin{quotation}\noindent
The equivalence between the Chern-Simons gauge theory on a
three-dimensional manifold with boundary and the WZNW model on the boundary is
established in a simple and general way using the BRST symmetry. Our approach is based on
restoring gauge invariance of the Chern-Simons
theory in the presence of a boundary. This gives a correspondence to the
WZNW model that does not require solving any constraints, fixing the gauge
or specifying boundary conditions. 

\end{quotation}

\vs{1cm}
{\em Submitted for publication to Physics Letters B}

\np
\setcounter{page}{1}
In a seminal paper \cite{wit1} on the 2+1 dimensional
Yang-Mills theory with an action consisting purely of a Chern-Simons term
("Chern-Simons gauge theory"), Witten showed that the Chern-Simons theory on a
compact surface was
intimately connected with two-dimensional conformal field theory and more
precisely to WZNW models with compact groups. The connection was stated in the
somewhat abstract form that the physical Hilbert spaces obtained by
quantization in 2+1 dimensions can be interpreted as the spaces of conformal
blocks in 1+1 dimensions. A more concrete connection was also suggested by
considering the Chern-Simons theory on a manifold with boundary. This
connection was further elaborated in \cite{emss} and \cite{ms}, generalizing the
connection to gauged WZNW models with compact groups.

Although the treatment in \cite{wit1}-\cite{emss} for a manifold with boundary
is explicit it is not entirely satisfactory for several reasons. Firstly, the
action on a manifold with boundary is not gauge invariant. A gauge
transformation yields boundary terms. This implies that upon quantization there
is no symmetry that restricts possible quantum corrections. However, the fact
that the gauge invariance is broken at the boundary is precisely what is used
in \cite{ms}, \cite{emss} to show that Chern-Simons action reduces to a chiral
WZNW action in a particular partial gauge. Thus, it seems that connection
between the Chern-Simons theory and the WZNW model can only be made if gauge
invariance is broken at the boundary and for a particular partial gauge.
Related to this is
the fact that in making the connection to the WZNW model one does not encounter
the standard chiral WZNW Hamiltonian, but in fact one that is zero \cite{ms}.
This is a consequence of the way the connection is made by solving the
constraints. It really means that we still have an action not suitable for
quantization since it is still not gauge fixed. 

Secondly, as there is no
gauge invariance on the boundary, one may even classically modify the
Chern-Simons action with boundary terms. Different boundary terms will yield
different boundary actions and different actions for the
conformal field theory. An argument due to Regge and Teitelboim \cite{rt}
may be invoked as a guiding principle for possible boundary terms. For our
particular example this is done in \cite{ban}. Notice however that the situation
in \cite{rt} is quite different from ours and we will see that in treating our
model we use a completely different principle.

These objections to the explicit correspondence should be
contrasted to the general connection involving the Hilbert space of the 2+1
dimensional theory and conformal blocks mentioned above. Although formulated for the case
of a compact
manifold without boundary and for a compact group, it does not refer
to any particular gauge or any particular choice of boundary condition. It is
natural to ask whether or not a more explicit formulation exists, which still
is general with no reference to a particular gauge or choice of boundary
condition.  Some progress towards such a formulation have been achieved see e.g. \cite{bbgs}, \cite{par}, but as far as we know there is no formulation which 
gives the connection between the CS and WZNW theories in a completely general and gauge
invariant way and at the level of actions, and which does not depend on any particular boundary conditions. 
The objective of the present work is to provide such a formulation.

We will see that the correspondence may be formulated in terms of a BRST
symmetry. Invariance with respect to the BRST transformations will imply that the two
theories - Chern-Simons and WZNW - are equivalent. Particular forms of the
action may then be found by different choices of the BRST invariant gauge
fixing terms. The crucial step in our approach is to restore gauge invariance of the CS
theory in
the presence of a boundary. This will be achieved by adding new dynamical degrees of
freedom at the boundary, namely the WZNW degrees of freedom. In spirit, our approach
is similar to that of Dirac \cite{dir}, where he advocated the introduction of new
degrees of freedom describing the initial quantization surface and corresponding
constraints to make the theory reparametrization invariant (see also \cite{rt}).
An even closer resemblence is to the work on surface terms for Yang-Mills theories
\cite{gsw},\cite{wad} in connection with non-Abelian monopole solutions. 

An important outcome of our treatment is
that the action is completely fixed up to BRST exact terms. It is no longer
possible to add boundary terms to the action without spoiling the BRST symmetry (with the 
exception of non-trivial BRST invariant terms, if they exist).
 Thus the apparent arbitrariness of adding boundary
terms in the original formulation of the CS theory (which however can be partially reduced by
appropriate boundary conditions) has been completely eliminated.

The use of the BRST symmetry in connection with the CS theory first appeared in \cite{bc}. 
Their treatment is, however, quite different from 
ours as 
gauge invariance is not restored (it uses a particular gauge and no new dynamical
degrees of freedom are introduced at the boundary) and certain 
boundary conditions are assumed. Also the connection to the CS theory is not achieved on the
level of actions.

Recently, the interest in the Chern-Simons theory has grown substantially due
to the work of Maldacena \cite{mal} on the correspondence between string theory
on anti-de Sitter spaces and certain conformally invariant theories associated
with
the boundary. This correspondence was formulated as the so-called holographic
principle \cite{gkp},\cite{wit3} (ideas originally presented in \cite{thooft}, \cite{sus}).
As our work here deals with a special (and simple) case
of the holographic principle it may be of some interest that there exists a
simple and elegant way to implement this principle. However, due to the
simplicity of the model with no propagating degrees of freedom in the bulk, it
is not clear that this may be of use for a more non-trivial example.  

The motivation of our work does not primarily come from AdS/CFT correspondence
in itself, but rather from our desire to understand the calculations
of the entropy of the 2+1 dimensional black hole presented in \cite{car1} 
(see also \cite{car2}). In
particular, we wanted to understand the problem of choice of boundary terms in
connection with the BTZ \cite{btz} black hole. Our results in connection with
this problem will be presented elsewhere \cite{fh}.

We start by briefly recalling the correspondence between Chern-Simons
theory on a manifold with boundary and WZNW models as formulated in
\cite{wit1}-\cite{emss}. The Chern-Simons action is
\be I_{CS}=-{k\over 2}\int_{\bf{R}\times \Sigma}Tr(A\wedge
dA+\frac{2}{3}A\wedge
A\wedge A).\e{1}
The action is invariant under the gauge transformation $\del A=d\ep + [A,\ep]$
if $\Sigma$ has no boundary. If $\Sigma$ has a boundary the variation yields a
boundary term 
\be \del I_{CS}=-{k\over 2}\int_{\bf{R}\times \partial\Sigma}Tr(\ep dA).\e{2}
The invariance of the action may be achieved by requiring $\ep=0$ on the
boundary. This is equivalent to saying that the full gauge group $G$ is reduced
to all gauge transformations $G_1$, that are one on the boundary. The variation
of $A_0$ yields a constraint ($i,j=1,2$)
\be \ep^{ij}F_{aij}\approx 0, \ a=1,\ldots n.\e{3}
This constraint implies that the vector field is pure gauge i.e. $A_i=-\partial
_i UU^{-1}$, for some map $U: \ \Si\rightarrow G$. The gauge invariance implies
that we have an equivalence relation $U\sim VU$ for any $V$ that is one on the
boundary. This implies that only the restriction of $U$ to the boundary is
relevant. If we use $A_i=-\partial _i UU^{-1}$,
$i=1,2$, and fix the gauge partially by $A^0=0$, the Chern-Simons action
reduces to 
\be I_{CS}=-{k\over 2}\left(\int_{\bf{R}\times
\partial\Sigma}Tr(U^{-1}\partial_\phi
UU^{-1}\partial_t U+\frac{1}{3}\int_{\bf{R}\times
\Sigma}
Tr(U^{-1}dUU^{-1}dUU^{-1}dU)\right).
\e{4}
Here we have chosen $\Si=D$, a disc of radius $R$ with a radial coordinate
$x^1=r$ and an angular coordinate
$x^2=\phi$. We
will for convenience make this choice for the remaining part of the
paper. As remarked above, the Hamiltonian corresponding to this action is
easily checked to be zero. Furthermore, the action corresponds to a chiral WZNW
action, where the current $J(z)$ in the conventional form of the WZNW theory here
is identified with $A_\phi=-\partial_\phi UU^{-1}$, as can be seen from
the form of the action \r{4}. The Dirac brackets of
these currents give  that the current satisfies an affine Lie
algebra of level $k$ \cite{wit2}. If we add to the action boundary terms e.g.
a term
$-{k\over 2}\int_{\bf{R}\times
\partial\Sigma}A_\mu A_\nu C^{\mu\nu}$, for some matrix $C^{\mu\nu}$
($\mu,\nu=0,1,2$), then inserting again
$A_i=-\partial _i
UU^{-1}$ leads to modifications of \r{4} by boundary terms, which in turn
implies that the resulting action is not the conventional chiral WZNW action. A
boundary term of this form (for a particular $C^{\mu\nu}$) was e.g. suggested
in
\cite{car1} for the BTZ black hole.

We now proceed to find an alternative way of realizing the connection between
the two theories. The Hamiltonian corresponding to the original action \r{1}
is
\be H_{CS}=2\int_Dd^2x\psi_aA^a_0-\frac{k}{2}\int_{\partial D}d\phi A^a_2A_{a0}.\e{5}
Let us neglect the surface term for the moment. Then the primary and secondary constraints are
\be P_a\approx 0 \mbox{ and } \psi_a\equiv
\frac{k}{4}\ep^{ij}F_{aij}\approx 0.\e{6}
Here $P_a$ is the momentum conjugate to $A_0^a$. Using the Poisson bracket (PB)
\newline \mbox{$\{A_2^a(x),\\ A_{b1}(x')\}=\frac{2}{k}\delta^a_b\delta^2(x-x') $} one
finds
\be
\{\psi_a(x),\psi_b(x')\}&=&-f_{ab}^c\psi_c\delta^2(x-x')\nonumber\\&&-k(
f_{ab}^cA_{c2}\delta^2(x-x')-\eta_{ab}\partial_\phi \delta^2(x-x'))\delta
(x^1-R).\e{7}
As the appearance of the boundary terms involving the delta function $\delta
(x^1-R)$ may seem surprising and is important in what follows, we will explain
the calculation in a little more detail. In computing the PB one needs to use
certain delta-function identities. These identities may be derived using test
functions, that are usually assumed to be zero on the boundary. Then the
constraint algebra may be shown to close. The extra terms on the right hand
side of eq.\r{7} appear if one generalizes the delta-function identities to hold
on test functions that are arbitrary on the boundary\footnote{The practical
way of making all computations here and below is in fact to use
$\psi_\la=\int_Dd^2x(\psi_a(x)\la^a(x))$ for some $L_1$ functions on D and
compute
$\{\psi_\la,\psi_{\la^\prime}\}$. An alternative way
of
computing the PB is to introduce a complete set of functions on $D$ and then by
using the PB of the modes with respect to this set, the PB can be computed yielding
the same result.}. The breakdown of a first class algebra due to a boundary was noted
previously in \cite{wad} for Yang-Mills theory, in \cite{sol1} for gravity in
the Ashtekar formalism, and in \cite{par} for the present case. In \cite{ber} and 
\cite{sol2} boundary
corrections for Poisson brackets in general are discussed.

The appearance of the extra terms in eq.\r{7} implies that the constraints are
not first class on the boundary i.e. gauge invariance is broken by boundary terms. 
This is of
course connected to the fact that the action is not gauge invariant due to
boundary contributions. It implies that a gauge may not be fixed on all of D.
The breaking of gauge invariance occurs only at the boundary. If $x,x'$ are
interior points, then the extra terms in eq.\r{7} are zero and the constraints
are first class. Further implications of the breakdown of gauge invariance is that the
time evolution of the secondary constraints yields additional constraints. 

In the careful computation of eq.\r{7} lies also the hint on how to
proceed. It is well-known that second class constraints may be converted into
first class constraints by adding new degrees of freedom, whereby restoring the
gauge symmetry of the theory. This will be the strategy in the following.
Firstly one notices that 
\mbox{$\tilde{\psi}_a(x)\equiv \psi_a(x) -\frac{k}{2}A_{a2}(x)\del(x^1-R) $}
satisfies
$\{\tilde{\psi}_a(x),\tilde{\psi}_b(x')\}=-f_{ab}^c\tilde{\psi}_c\delta^2(x-x')
-k\eta_{ab}\partial_\phi \delta^2(x-x')\delta (x^1-R)$, so that the
modification due to boundary terms are field independent. If we
define a level $k$ WZNW current $J_a(t,\phi)$ (we will surpress the $t$-dependence
henceforth) satisfying 
\be \{J_a(\phi),J_b(\phi')
\}=-f^c_{ab}J_c(\phi)\del(\phi-\phi')+k\eta_{ab}\partial_\phi\del(
\phi-\phi'),\e{8}
then 
\be \psi'_a(x)\equiv \psi_a(x)
-\frac{k}{2}A_{a2}(x)\del(x^1-R)-J_a(\phi)\del(x^1-R)\e{9}
satisfies the closed algebra
$\{\psi'_a(x),\psi'_b(x')\}=-f_{ab}^c\psi'_c\delta^2(x-x')$. Having first class
constraints we now define a Hamiltonian 
\be H'=2\int_D d^2x A_0^a\psi'_a=H_{CS}+\int_{\partial D}d\phi
(-\frac{k}{2}A_0^aA_{a2}+2J_aA_0^a).\e{10}
This Hamiltonian is automatically gauge invariant even at the boundary. Notice
that gauge invariance forces a modification of the Chern-Simons action by a
{\em unique} boundary term in the vector fields. Let us pause and comment on
the constraints of the new theory. These are given
by $\psi'_a\approx 0$. Examining eq.\r{9} we see that these constraints consist
of two
pieces, a bulk and a boundary part. The boundary part enters much like a source
term. For any point in the interior of $D$ we simply have the original
constraints $\ep^{ij}F_{aij}\approx 0$. If we consider field configurations
$A^i_a$
that are continuous and have continuos spatial derivatives at the boundary, then
they will also satisfy $\ep^{ij}F_{aij}\approx 0$ on $\partial D$. In this case
the constraint implies
that $A_{a2}(\phi)\approx -\frac{2}{k}J_a(\phi)$ on $\partial D$. Thus, we recover the
connection between WZNW currents and $A_{2}^a$ mentioned above in the original
formulation. For field configurations
that
are not smooth at the boundary this relation does
not hold. It is not clear to us what the implications of this generalization
mean.

A
simple counting of the degrees of freedom shows that the additional degrees of
freedom associated with the currents $J_a$, which are $n$ phase space degrees
of freedom at every space point on the boundary, compensate exactly that the
$n$ constraints are transformed from second class
to first class constraints on the boundary. Let us now show in more detail that
the theory defined by this Hamiltonian is equivalent to the Chern-Simons theory
defined by the action \r{1} and a specific boundary term added to it. 

First we
impose the partial gauge $A_0^a\approx 0$. Then
$H_{CS}\approx H'\approx 0$ so that the resulting actions are identical. It remains to
show
that we can break the gauge invariance by imposing gauge fixing constraints on
the boundary that eliminate the new degrees of freedom. We consider a gauge
constraint
$J_a(\phi)\delta(x^1-R)\approx 0$. To see that this fixes the gauge at the
boundary consider its gauge variation with parameter $\la_a$
\be
\{\int_Dd^2x\psi'_b(x)\la^b(x),J_a(\phi')\delta(x^{\prime 1}-R)\}\approx
-\frac{k}{2}
\partial_\phi^\prime\la_a(R,\phi').\e{10a}
Equating this to zero, we find $\partial_\phi^\prime\la_a(R,\phi')= 0$, from
which we conclude that the gauge is fixed at the boundary apart from the 
zero mode part of $\psi'_a$. Neglecting this detail for the moment,
the gauge fixing constraint eliminates the WZNW current degrees of freedom
reducing
the constraints $\psi'_a\approx 0$ to $\tilde{\psi_a}\approx 0$, which are second
class on
the boundary. Thus we are back to the original Chern-Simons theory (with a
specific boundary term). 

Let us now comment on the zero mode part. The gauge
fixing of the zero mode part, $\int d\phi J_a(\phi)\approx 0$, is not a valid
gauge
choice. We may notice that in fact the $\phi$-independent part of the constraint 
generators
$\psi'_a$ satisfy a closed algebra. Thus, we have really introduced slightly
more degrees of freedom than necessary. This may be changed by subtracting the
zero mode of $J_a$ from $\psi'_a$. Then the $\phi$-independent
part of $A_2^a$ will play the
r\^ole of the zero mode of the current. In the following we will for simplicity
not concern ourself with this subtlety and leave $\psi'_a$ unaltered. 

We have already seen that our theory reduces to the Chern-Simons theory for a
particular choice of partial gauge. We will now show that for another choice
of gauge we may eliminate the vector field degrees of freedom leaving us with
only the WZNW degrees of freedom, thus establishing classically the equivalence
of the two theories. Imposing the gauge constraints
$A_0^a\approx A_1^a\approx 0$, it is easily checked that this fixes the gauge
both in the interior of $D$ and at its boundary. This eliminates all vector
field degrees of freedom (as $A_2^a$ is conjugate to $A_1^a$). The Dirac
bracket is, therefore, only non-zero
for
brackets containing the WZNW degrees of freedom and, therefore, we have found
the promised
reduction. Possible forms for the Hamiltonian will be discussed below.

We have established the connection between the Chern-Simons theory and the
chiral WZNW model by imposing different gauges. A more fundamental way of
manifesting the equivalence is to introduce a BRST charge. As the constraints
are first class this is straightforward using the BFV formalism \cite{fv},
\cite{bv}. We have
\be
\Omega\equiv &
\int_Dd^2x[\frac{k}{4}\ep^{ij}F_{aij}c^a+\frac{1}{2}f^c_{ab}b_cc^ac^b+P_a\bar{b
}^a]\nonumber\\
& -\int_{\partial D}d\phi[(\frac{k}{2}A_{a2}+J_a)c^a].\e{11}
We note that the BRST charge consists of two distinct parts. In the first
line we have a bulk part and in the second a boundary part. The ghosts
$c^a,\bar{b}^a$ and their momenta $b_a,\bar{c}_a$ satisfy the usual PB. It is
easily checked that $\{\Omega, \Omega\}=0$. A BRST invariant Hamiltonian may
now
be constructed in the standard way
\be H_{tot}=\{\Omega,\chi\}\e{12}
for some gauge fermion $\chi$. We will now discuss some possible choices of
$\chi$. \vs{5mm}\\
 (i) $\chi=\int_Dd^2 x\left(b_aA^a_0+\bar{c}_a(\dot{A}^{a}_0+\chi^{\prime
a}[A])\right) + \int_{\partial D}d\phi \bar{c}_a J^{a},$\vs{3mm}\\
where $\chi^{\prime a}$ is some gauge fixing functional of the vector fields.
Inserting this into \r{12} yields a Hamiltonian
\be 
H_{tot}=&\int_Dd^2x[P_a\dot{A}^a_0+\bar{b}_a\dot{\bar{c}}^a+\psi^{tot}_aA_0^a+P
_a\chi^{\prime a}+c^a\{\psi'_a,\chi^{\prime b}\}\bar{c}_b]\nonumber\\
&+\int_{\partial D}d\phi[f^c_{ab}J_c\bar{c}^ac^b+P_aJ^a].\e{13}
Here $\psi^{tot}_a\equiv \{\Omega,b_a\}$. $P_a$ acts as Lagrange
multipliers for the gauge fixing constraints. The eq. of motion for $A_0^a$
give $\chi'_a+J_a\delta(x^1-R)=0$, which for smooth vector fields at the
boundary yields $\chi'_a=0$ on $D$ and $J_a=0$ on $\partial D$. This
Hamiltonian, therefore, corresponds to eliminating the WZNW degrees of
freedom.\vs{5mm}\\
(ii) $\chi=\int_Dd^2
x\left(b_aA^a_0+\bar{c}_a(\dot{A}^{a}_0+A_1^a)\right) + \int_{\partial D}d\phi \bar{c}_a
\left(J^{a}-\frac{\alpha}{2}P^{a}\right).$\vs{3mm}\\
This choice gives a Hamiltonian
\be
H_{tot}=&\int_Dd^2x[P_a\dot{A}^a_0+\bar{b}_a\dot{\bar{c}}^a+\psi^{tot}_aA_0^a+P
_aA^{a}_1+c^a\{\psi'_a,A^{b}_1\}\bar{c}_b]\nonumber\\
&+\int_{\partial
D}d\phi[f^c_{ab}J_c\bar{c}^ac^b+P_a(J^a-\frac{\al}{2}P^a)].\e{14}
Here the eq. of motion for $P_a$ imply
$A_1^a+(J^a-\al P^a)\delta(x^1-R)=0$, which for smooth fields give $A_1^a=0$ in
$D$ and $J^a-\al P^a=0$ on $\partial D$. This choice corresponds to eliminating
the vector field at the boundary. Inserting the second equation into the
Hamiltonian gives
\be H_{tot}=\int_{\partial D}d\phi
[\frac{1}{2\al}J^aJ_a+f^c_{ab}J_c\bar{c}^ac^b] + H_{bulk}.\e{15}
The first term is precisely the Sugawara Hamiltonian (for an appropriate choice
of $\al$). The second term is a ghost correction at the boundary. The third
term is a bulk Hamiltonian. Since the degrees of freedom are completely
eliminated in the bulk, this term will not contribute to any Dirac brackets. In
fact, we may set this term to zero and still have a BRST invariant Hamiltonian.

Other choices of gauge fermions are of course possible e.g. one which makes the
Lagrange multiplier field $A_0^a$ dynamical. We will, however, not discuss this
further. Instead let us end with some concluding remarks.

We have demonstrated the classical equivalence between the Chern-Simons  gauge
theory and the chiral WZNW model. It should be emphasized that the equivalence
between the bulk and boundary theories follows without any specifications of
gauge or boundary conditions. It is the gauge invariance or more generally
the BRST invariance that makes the different choices of Hamiltonian physically
equivalent. The quantum equivalence is ensured once the nilpotency of the BRST
operator is established and  non-trivial BRST invariant states are found. Thus,
for this particular example one may state that the holographic principle is
translated into a statement about BRST symmetry. 

Our considerations have for simplicity been for the disc, but they may easily
be generalized to other cases. Then the delta function $\delta(x^1-R)$ is
replaced by a generalized delta function $\delta(x\in \partial \Si)$, which is
defined by $\int_\Si[f(x)\delta(x\in \partial \Si)]=\int_{\partial \Si}f(x)$. Note also
that it is irrelevant where the boundary actually is
i.e. we have the freedom to move it at will. This is a freedom that follows from our
formulation as it does not rely on any particular boundary conditions. For the
case of the black hole it will imply that the calculation of the entropy will
be (almost) independent of where we actually put the boundary. We can choose
the horizon, but just as well another surface. We will discuss this in more
detail in \cite{fh}. \vs{10mm}\\
{\bf Acknowledgement:} We would like to thank Steve Carlip for enlightening
discussions concerning his work. We are also indepted to Ioannis Bakas for pointing out the
work in \cite{gsw} and \cite{wad}.

\end{document}